\documentclass{aa}

\newcommand{\C}{\v{C}erenkov }
\newcommand{\etal}{{et al. }}

\begin{document}


\title{A study of Tycho's SNR at TeV energies with the HEGRA CT-System}

\author{F.A. Aharonian\inst{1}, A.G.~Akhperjanian\inst{7}, J.A.~Barrio\inst{2,3}, K.~Bernl\"ohr\inst{1,}, H.~B\"orst\inst{5},
        H.~Bojahr\inst{6}, O.~Bolz\inst{1}, J.L.~Contreras\inst{3}, J.~Cortina\inst{2}, 
        S.~Denninghoff\inst{2}, V.~Fonseca\inst{3}, J.C.~Gonzalez\inst{3}, N.~G\"otting\inst{4}, G.~Heinzelmann\inst{4},
        G.~Hermann\inst{1}, A.~Heusler\inst{1}, W.~Hofmann\inst{1},
        D.~Horns\inst{4}, A.~Ibarra\inst{3}, I.~Jung\inst{1}, R.~Kankanyan\inst{1,7}, M.~Kestel\inst{2},
        J.~Kettler\inst{1}, A.~Kohnle\inst{1}, A.~Konopelko\inst{1},
        H.~Kornmeyer\inst{2}, D.~Kranich\inst{2}, H.~Krawczynski\inst{1}$^\dag$, H.~Lampeitl\inst{1},
        E.~Lorenz\inst{2}, F.~Lucarelli\inst{3}, N.~Magnussen\inst{6}, O.~Mang\inst{5}, H.~Meyer\inst{6}, R.~Mirzoyan\inst{2}, 
        A.~Moralejo\inst{3}, 
        L.~Padilla\inst{3}, M.~Panter\inst{1}, R.~Plaga\inst{2},
        A.~Plyasheshnikov\inst{1}$^\S$, J.~Prahl\inst{4}, G.~P\"uhlhofer\inst{1}, G.~Rauterberg\inst{5}, A.~R\"ohring\inst{4},
        W.~Rhode\inst{6}, G.P.~Rowell\inst{1}, V.~Sahakian\inst{7}, M.~Samorski\inst{5}, M.~Schilling\inst{5},
        F.~Schr\"oder\inst{6}, W.~Stamm\inst{5}, M.~Tluczykont\inst{4}, H.J.~V\"olk\inst{1}, C.~Wiedner\inst{1},
        W.~Wittek\inst{2}}

\institute{Max Planck Institut f\"ur Kernphysik,
           Postfach 103980, D-69029 Heidelberg, Germany \and
           Max Planck Institut f\"ur Physik, F\"ohringer Ring
           6, D-80805 M\"unchen, Germany \and
           Universidad Complutense, Facultad de Ciencias
           F\'{\i}sicas, Ciudad Universitaria, E-28040 Madrid, Spain 
           \and
           Universit\"at Hamburg, II. Institut f\"ur
           Experimentalphysik, Luruper Chaussee 149,
           D-22761 Hamburg, Germany \and
           Universit\"at Kiel, Institut f\"ur Experimentelle und Angewandte Physik,
           Leibnizstra{\ss}e 15-19, D-24118 Kiel, Germany\and
           Universit\"at Wuppertal, Fachbereich Physik,
           Gau{\ss}str.20, D-42097 Wuppertal, Germany \and
           Yerevan Physics Institute, Alikhanian Br. 2, 375036 Yerevan, 
           Armenia\\
           \hspace*{-4.04mm} $\dag\,$ \emph{Present address}: Astronomy Dept., Yale University,
           P.O. Box 208101, New Haven, CT 06520-8101 USA\\
           \hspace*{-4.04mm} $^\S\,$ On leave from  
           Altai State University, Dimitrov Street 66, 656099 Barnaul, Russia}

\offprints{G.P. Rowell, \email{Gavin.Rowell@mpi-hd.mpg.de}}

\date{Received / Accepted}

\abstract{Tycho's supernova remnant (SNR) was observed during 1997 and 1998 with the
 HEGRA \C Telescope System in a search for gamma-ray emission at energies above
 $\sim$ 1 TeV. An analysis of these data, $\sim$65 hours in total, resulted in
 no evidence for TeV gamma-ray emission. The 3$\sigma$ upper limit to the 
 gamma-ray flux ($>$1 TeV) from Tycho is estimated at $5.78\times 10^{-13}$ photons cm$^{-2}$
 s$^{-1}$, or 33 milli-Crab. We interpret our upper limit within the framework of the following scenarios:
 (1) that the observed hard X-ray tail is due to synchrotron emission. A lower limit on 
 the magnetic field within Tycho may be estimated $B\geq22$ $\mu$G, assuming that the RXTE-detected
 X-rays were due to synchrotron emission. However, using results from a detailed model of the ASCA emission,
 a more conservative lower limit $B\geq6$ $\mu$G is derived. 
 (2) the hadronic model of Drury, Aharonian \& V\"olk, and (3) the more recent 
 time-dependent kinetic theory of Berezhko \& V\"olk. 
 Our upper limit lies within the range of predicted values of both hadronic models, according to uncertainties
 in physical parameters of Tycho, and shock acceleration details. In the latter case, the model was scaled 
 to suit the parameters of Tycho and re-normalised to account for a simplification of the original model. 
 We find that we cannot rule out Tycho as a potential contributor at an average level to the Galactic 
 cosmic-ray flux.
 }

\titlerunning{Tycho's SNR at TeV energies}
\authorrunning{F.A. Aharonian et al.}

\maketitle

\section{Introduction}
 \label{sec:intro}

The search for gamma-ray emission of TeV energies from supernova remnants (SNRs) in recent years
is motivated by the need to explain the origin of Galactic cosmic-rays (CR). SNRs are
long-believed primarily responsible for the Galactic CR population, matching the CR
energetics and spectral index (V\"olk \cite{Volk:1},
Baring \cite{Baring:1}). The production of
$\gamma$-rays in SNRs is thought to result from the interaction of 
shock-accelerated particles (hadrons and electrons) with the interstellar medium (ISM) and
local soft photon fields. Being relatively unattenuated over long distances and
preserving the production site directionality, gamma radiation is one of the
most accessible tracers of CR acceleration in the universe.

TeV gamma radiation is primarily expected from two channels (1) Collisions 
of hadronic CRs, producing gamma-rays via $\pi^\circ$ decay and (2) CR electrons up-scattering
soft photons via the inverse Compton (IC) process, and CR electron collisions via Bremsstrahlung.
Detailed modelling of SNR environments has revealed distinct spectral features in the GeV/TeV regime for 
these processes, and in combination with radio and X-ray observations, those in the TeV regime
are deemed vital in establishing SNRs as sites of CR production (Drury, Aharonian \& V\"olk \etal \cite{Drury:1}, 
Naito \& Takahara \cite{Naito:1}, Berezhko \& V\"olk \cite{Berezhko:1}, Baring \etal \cite{Baring:2}).

So far, evidence for TeV emission has come from CANGAROO observations of two southern hemisphere SNRs,
SN 1006, and SNR RX J1713.7-3946 (Tanimori \etal \cite{Tanimori:1}, Muraishi \etal \cite{Muraishi:1}), 
and the HEGRA \C Telescope CT-System (High Energy Gamma Ray Astronomy \C Telescope) 
after deep observations of Cas-A (P\"uhlhofer \etal \cite{Puhlhofer:1}, 
Aharonian \etal \cite{Aharonian:1}).
Upper limits are reported for those other promising SNR candidates observed so far in both the TeV and PeV regimes  
(Buckley \etal \cite{Buckley:1}, Prosch \etal \cite{Prosch:1}, Goret \etal \cite{Goret:1}, Allen \etal \cite{Allen:1}, 
Rowell \etal \cite{Rowell:1}),
including Tycho's SNR, for which the Whipple Collaboration obtained 14.5 hours data. The CANGAROO and HEGRA detections
might be interpreted in the framework of the SNR as a source of multi-TeV CR electrons by virtue of a strong non-thermal
tail in their X-ray spectra above 1 keV (Koyama \etal \cite{Koyama:1,Koyama:2}, Allen \etal \cite{Allen:1}). On the other hand 
as argued by Aharonian \& Atoyan (\cite{Aharonian:6}) and Berezkho, Ksenofontov \& Petukhov (\cite{Berezhko:3}) for SN 1006, and
by Atoyan \etal (\cite{Atoyan:1}) for Cas-A, the TeV results do not rule out the hadron channel. 
It is nevertheless less clear at the moment 
as to the location of CR hadron accelerators in our galaxy.  

Tycho's SNR (G120.1$+$1.4, 3C 10) is one of the most intensely studied SNRs. It is an archetypal shell-type
(radio \& X-Ray) SNR, formed most likely from a type Ia supernova (SN 1572), and has expanded at $\sim$0.1\% yr$^{-1}$ to a radius 
$\sim 4^\prime$ (Katz-Stone \etal \cite{Katz-Stone:1} and references therein). One estimate of distance is put at 2.2 kpc 
(Albinson \etal 1986 and references therein) based on proper motion studies, absorption against field stars and the fact
that Tycho appears embedded in the Perseus arm of the Milky Way. A higher distance estimate of 4.5 kpc (Schwarz \etal 
\cite{Schwarz:1}) is derived from a model of the HI spectrum and number of HI absorbing features in the region. The radio synchrotron
emission (20 \& 90 cm) shows variation in photon index that appears correlated with edge filaments, perhaps tracing
regions of enhanced magnetic field and particle acceleration (Katz-Stone \etal \cite{Katz-Stone:1}). Studies at the HI 21 cm line
indicate expansion on the eastern side of Tycho is slowed by a region of higher density (160--325 cm$^{-3}$, Reynoso \etal
\cite{Reynoso:1}), suggesting that Tycho's SNR may not be expanding into such a homogeneous region as earlier believed.
The X-ray emission shows very strong line features and overall, is well fit by thermal bremsstrahlung components
(Fink \etal \cite{Fink:1}, Petre \etal \cite{Petre:1}). A power law however is necessary to fit a hard X-ray tail.
This presumably non-thermal tail above 1 keV suggests that Tycho may a source of electrons up to $\sim$100 TeV, thereby 
joining the growing number of SNRs that exhibit this feature (see Petre \etal \cite{Petre:1} for a review).
  
\section{Observations}
The observations of Tycho's SNR were made using the HEGRA CT-System 
incorporating, at the time of data taking (1997 \& 1998), four identical imaging air \C telescopes\footnote{Since 1999, the full 
five telescopes have been in operation, CT2 being the most recent telescope brought online.}(CT3-6). 
The telescopes are operated in coincidence to achieve a stereoscopic view of \C light air showers induced by $\gamma$ and 
CR primaries. The CT-System is situated on the Roque de los Muchachos at La Palma (2200 m asl, $28^\circ \, 45^\prime$ N 
$17^\circ \, 54^\prime$ W) and at the zenith angles of these data, operates at an energy threshold for $\gamma$-ray primaries of 
$\sim$1 TeV (see Konopelko \etal \cite{Konopelko:1} and references therein). 

\begin{table}
\begin{center}
 \begin{tabular}{lrrr}\\ \hline \hline \\
                      & 1997     & 1998     & Total \\ \hline \\
  Obs. time (hrs)     & 19.7     & 44.9     & 64.6  \\
  Selected runs       & 50       & 115      & 165(82\%) \\
  Configuration       & CT3-6    & CT3-6    &       \\
  Zenith angle range  & 32-41$^\circ$ & 32-45$^\circ$ & \\ 
  CR Event Rate (Hz)  & 8.4 (10.3)           & 8.5     & 8.5  \\ \hline \hline
 \end{tabular}
 \caption{Summary of Tycho Observations by the HEGRA CT-System. The CR event rate is after the {\em size} cut of 40 photoelectrons
          and the CR rate in brackets for 1997 is prior to a cut of 10 photoelectrons on second brightest image pixel.}
 \label{tab:tycho_data}
\end{center}
\end{table}
The Tycho data were accumulated in Jul-Sep 1997 and Aug-Dec 1998 in 20$-$30 minute runs utilising the {\em wobble} mode of
observation. In this mode, the source is offset $\pm 0.5^\circ$ in declination in alternating runs, enabling the background or OFF
source data to be estimated from the opposing position (or a series of positions) 1.0$^\circ$ away. Table~\ref{tab:tycho_data} gives 
a summary of these observations.

Approximately 20\% of all runs were rejected, due primarily to weather effects. Bad runs were identified
according to the criteria that the image {\em width} from each triggered telescope and stereo trigger rate match their expected values
accordingly, within 6\% for {\em width} and 15\% for rate (representing levels $<4\sigma$ outside their Gaussian distribution means).
The bad runs actually comprised the non-Gaussian tails of these distributions. 
The trigger rate for CRs was significantly lower 
in 1998 compared to 1997 (by $\sim$25\%), attributed mainly to photomultiplier tube (PMT) fatigue and mirror degradation.
We accounted for this threshold difference by artifically
raising the trigger threshold of 1997 data to 1998 levels by applying a 10 photoelectron threshold to the {\em second}-brightest
pixel in the image, mimicking the hardware trigger threshold, at 8 photoelectrons. The effect of this software trigger is to simplify
analysis and in particular, the estimation of upper limits. It may be justified given the negligible effect on the $\gamma$-ray 
threshold of this cut (following the 40 photoelectron cut described below) and the fact that the 1997 observation time is $\sim$45\% 
that of 1998. The $\gamma$-ray energy threshold\footnote{The maximum differential trigger rate, for a number of photon indices.} 
for both years' data were found from simulations to be consistent within systematic errors (15\%) to a value of 1 TeV.

Event filtering follows the standard method in that images containing more than 40 photoelectrons ({\em size}) are selected for 
further analysis. The rejection of the CR background is achieved with cuts on the image {\em shape} and the location of it's
reconstructed source position. The reconstruction is described as ``Algorithm 1'' by Hofmann \etal (\cite{Hofmann:1}) and 
the shape cut used is the so-called {\em mean scaled width (msw)}. 
The {\em msw} is the average of the {\em scaled widths} from each accepted image. Each image is scaled according to it's expected
value ($\gamma$-ray hypothesis), dependent upon image {\em size}, zenith angle of observation and impact distance of the event from the 
CT-System centre. Following cuts on {\em msw} and $\theta$, the difference between the reconstructed and actual source
positions, the stereoscopic method permits unprecedented angular and energy 
resolution for $\gamma$-ray images on an event-by-event basis of $\sim 6^\prime$ and $\sim$20\% respectively 
(Aharonian \etal \cite{Aharonian:2}). 

\begin{figure}
 \vspace{8.5cm}
 \includegraphics{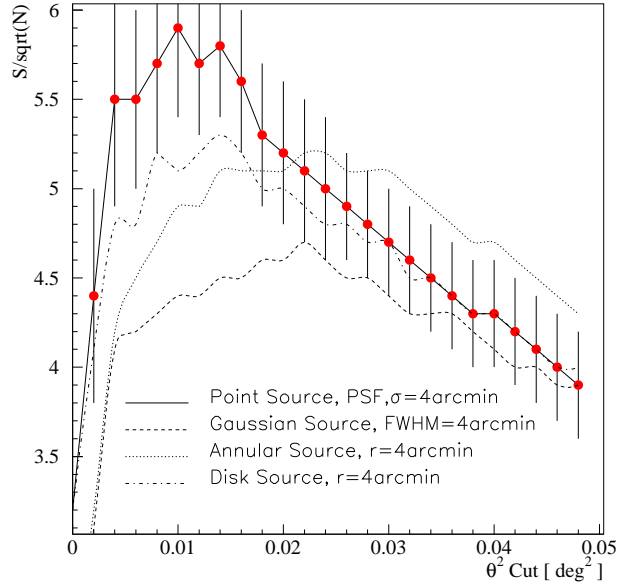}
 \caption{Signal to noise ratios (unnormalised curves) from a Monte Carlo for sources of various morphologies. The HEGRA CT-System
          point spread function (PSF, with $\sigma$ specified in the plot) is assumed 
          to be a Gaussian and based on measurements on the Crab (Aharonian \etal \cite{Aharonian:5}). Here 100000 background 
          events were sampled, reproducing roughly the statistics of the Tycho dataset, and the number of source events chosen to 
          reproduce a $\sim5\sigma$ maximum excess.}
 \label{fig:besttheta2}
\end{figure}

We use a combination of ``tight cuts'' in the shape (0.4$<${\em msw}$<$1.1) and direction 
($\theta^2<0.02$ deg$^2$) criteria to preferentially select $\gamma-$ray images against those from CR. Tight cuts are optimal 
in the search for weak signals in background dominated statistics (3$-$5$\sigma$ level), with the caveat that systematic effects may
become more important due to the cuts lying close to distribution tails. Tycho is a marginally extended source for the 
HEGRA CT-System, and thus the optimal $\theta^2$ cut will be slightly larger than that 
for a point source at $\sim$0.015 deg$^2$. A simple Monte Carlo, matching roughly the statistics of the background of our Tycho dataset,
was used to determine an optimal $\theta^2$ cut, assuming various morphologies for the SNR  with the results displayed in 
Fig.~\ref{fig:besttheta2}. 
Given the uncertainty in the SNR morphology (disk, Gaussian, annular or a combination), a $\theta^2$ cut of 0.02 deg$^2$ was 
deemed suitable.

\begin{figure}
 \vspace*{5cm}
 \includegraphics{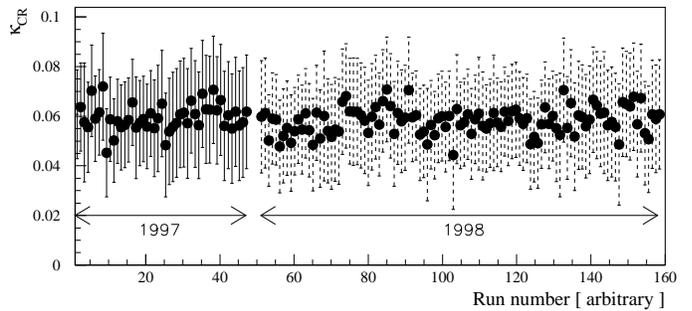}
 \label{fig:kappaCR}
 \caption{Cosmic ray acceptance ($\kappa_{\rm CR}$) of the {\em msw} cut on a run-by-run basis. The 1997 data were also subjected
         to a cut on the second-brightest pixel of 10 p.e.}
\end{figure}

A useful indicator of data quality is the CR acceptance ($\kappa_{\rm CR}$) of the {\em msw} cut on a run-by-run basis 
(Fig.~\ref{fig:kappaCR}).      
We can see that $\kappa_{\rm CR}$ is consistent over the entire dataset and moreover, the cut of 10 p.e on the second brightest pixel
applied to 1997 data introduced no significant change in $\kappa_{\rm CR}$.                                                             

\section{Results}
\begin{figure}
 \vspace{5cm}
 \includegraphics{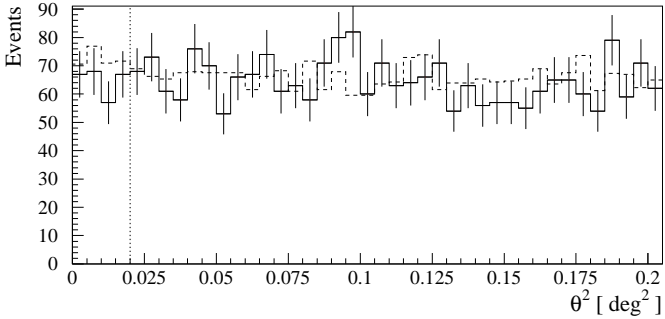}
 \caption{Distribution of $\theta^2$ for combined 1997 and 1998 observations of Tycho's SNR (ON source, solid line). The OFF source 
          data (dashed line) are taken from three positions in the field of view (see text) and scaled by a factor $\frac{1}{3}$. 
          The optimal cut is indicated by the dotted vertical line} 
 \label{fig:theta2}
\end{figure}

We evaluate the statistical significance of the post-cut excess using equation 9 of Li \& Ma (\cite{Li:1}). The background or OFF source
counts are taken from three control regions at position angles 180$^\circ$ \& $\pm 90^\circ$, on a circle 0.5$^\circ$ radius centred on
the tracking position, as the wobble mode of observation permits. A normalisation factor of $\frac{1}{3}$ is therefore used in the 
significance calculation. By using more than one
background region, effects of skynoise, camera response and zenith angle difference are averaged out further (on
top of that already afforded by the stereoscopic technique), reducing somewhat any
systematics expected from such properties, and moreover, reducing the statistical error in the background.
A distribution of $\theta^2$ for the combined data after application of the {\em msw} cut discussed above is 
presented in Fig.~\ref{fig:theta2}. 
We summarise further the results in Table~\ref{tab:excesses}. No significant excess was seen in either dataset, and the combined
excess represents a significance of $-$1.4$\sigma$. 
\begin{table}
 \begin{center}
 \begin{tabular}{lrrr}\\ \hline \hline \\
                            & 1997     & 1998     & Total  \\ \hline \\
   ON                       & 127      & 259      & 386    \\
   OFF  ($\Sigma^3_i b_i$)    & 383      & 871      & 1254   \\
   ON--OFF ($\sigma$)    & $-$0.7   & $-$1.6   & {\bf $-$1.4} \\
   3$\sigma$ UL (Crab)      &          &          & {\bf 0.033}  \\
   \multicolumn{3}{l}{3$\sigma$ UL ($F(>1 {\rm TeV}) \times 10^{-13}$ ph cm$^{-2}$s$^{-1}$)} & {\bf 5.78} \\ \hline \hline
 \end{tabular}
 \end{center} 
 \caption{Post-cut (after {\em msw} \& $\theta^2$ cuts) statistics of the Tycho data (significance calculated using 
          equation 9 of Li \& Ma (1983)) 
          taking as the background, the sum of positions $b_i$. The 3$\sigma$ upper limit (99.7\% UL) is expressed 
          in Crab and absolute units (see text).}
 \label{tab:excesses} 
\end{table}
We may express our upper limit in absolute units by using the TeV flux from the Crab Nebula, thereby taking 
advantage of the 
substantial simulation effort in deriving the absolute Crab flux (Aharonian \etal \cite{Aharonian:2}). Fluctuating the 
ON--OFF excess according to Helene (\cite{Helene:1}) and using the method described by Aharonian \etal \cite{Aharonian:3},
our 3$\sigma$ UL (99.7\%) corresponds to 33 milli-Crab\footnote{1 Crab: $F(>E\, {\rm TeV})=1.75\times 10^{-11}\left(\frac{E}{\rm 1\, TeV}\right)^{-1.59}$ 
ph cm$^{-2}$s$^{-1}$ (Aharonian \etal \cite{Aharonian:2})}. Here, those Crab data used to estimate the UL ($\sim$6 hours in total), were
subjected to an identical analysis as described above, and selected to match the zenith angle distribution as that for Tycho. Above the 
threshold energy of 1 TeV, our UL is $F(>1{\rm TeV})=5.78\times 10^{-13}$ ph cm$^{-2}$s$^{-1}$. 

Finally, we should point out that the method of Aharonian \etal \cite{Aharonian:3} 
(which downward fluctuates the Crab excess) 
to estimate the UL in Crab units was tested for accuracy with a simple Monte Carlo and found to overestimate a true 3$\sigma$ UL 
(for the numbers of Table~\ref{tab:excesses} by about 25\%. This overestimate is effectively canceled however by the negative 
bias of our $-$1.4$\sigma$ excess, an underestimate of $\sim$30\% relative to a zero excess result. We therefore may assume that our
3$\sigma$ UL represents closely a true estimate. Secondly, any integral UL or flux estimate quoted in absolute units will 
naturally depend somewhat on the difference in photon index between the Crab and Tycho. This uncertainty amounts 
to $\leq$30\% on the
UL estimation for a 20\% uncertainty in the values of 1 TeV and $-$1.59 respectively for energy and photon index.

\section{Inverse Compton Interpretation}
 \label{sec:ICint}

The evidence for a {\em non-thermal} tail (Fink \etal \cite{Fink:1}, GINGA data; 
Petre \etal \cite{Petre:1}, RXTE data) may suggest that 
Tycho's SNR is an accelerator of CR electrons to multi-TeV energies. In such a framework, the synchrotron process is assumed to 
account for these X-ray photons,
and the possibility of inverse Compton (IC) TeV $\gamma$-ray emission should be considered. However, one must be careful 
here since a power law fit is reasonable only at energies above $\sim$10 keV and, 
alternative physical explanations to the synchrotron scenario do exist (see for e.g. Asvarov \etal \cite{Asvarov:1}, 
Laming \cite{Laming:1}, and Tatischeff \etal \cite{Tatischeff:1}). 

\begin{figure}
 \vspace{12cm}
 \includegraphics{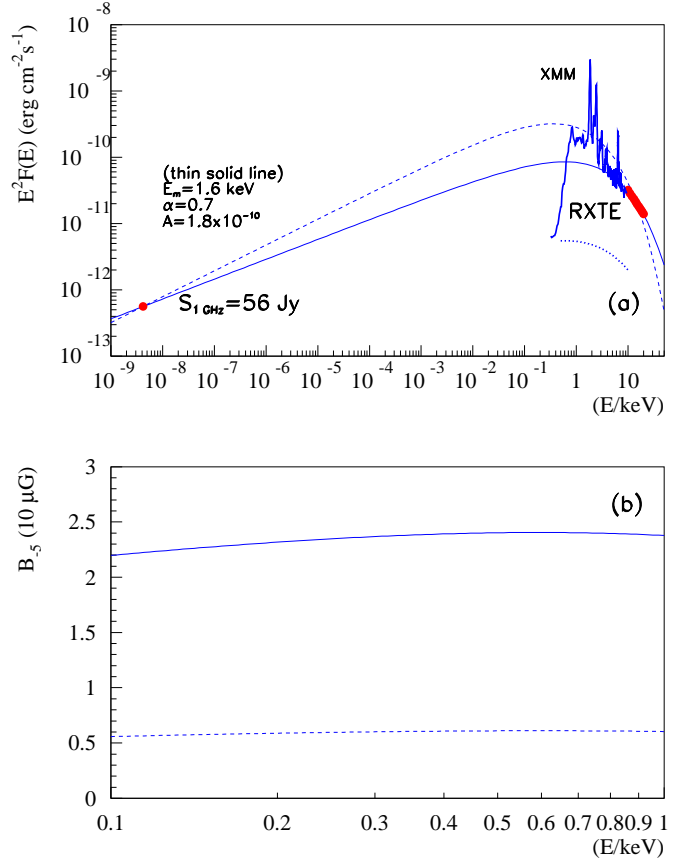}
 \caption{(a) Optimal fit to the RXTE spectrum and radio normalisation using Eq.~(\ref{eq:xrayfit}) (thin solid line). (dashed line) A 
     fit when fixing $\alpha$=0.61. (thick solid line) For comparative purposes a model fit to the XMM MOS 1 spectrum from 
     Decourchelle \etal (\cite{Decourchelle:1}) fig 1 is included. Since interstellar absorption 
     is not removed in this fit, the XMM flux below 1.0 keV is therefore a lower limit. (dotted line)
     Re-scaled RXTE fit to accommodate an upper limit to the non-thermal flux of Hwang \etal (\cite{Hwang:2}). 
    (b) (Solid line) Calculation of $B_{-5}$ (10 $\mu$G) of Eq.~(\ref{eq:ICandXray}) taking as $f_{\rm x}$ the fit given by the solid line of (a), the RXTE fit, and (dot-dashed line) $B_{-5}$ values obtained when using the
 scaled-down fit to a synchrotron spectrum given by the dotted line of (a) as suggested by 
 Hwang \etal (\cite{Hwang:2}).}
 \label{fig:fits}
\end{figure}

Assuming that the synchrotron model is valid for Tycho, we may use the 
direct relationship between the expected IC ($f_{\gamma}$) and X-ray energy fluxes\footnote{Energy flux: $E^2 F(E)$} ($f_{\rm x}$) arising 
from the {\em same} electrons:
\begin{equation}
 \frac{f_{\rm x}(\epsilon \,{\rm keV})}{f_\gamma(E\, {\rm TeV})} \sim 10 \left(\frac{B}{10^{-5}\, \rm G}\right)^2
 \label{eq:ICandXray}
\end{equation}  
to establish a condition on the magnetic field $B$ of the SNR. The mathematical caveat here is that the correct energy range in 
the TeV and keV regimes must be adhered to and that we assume that the emission regions of both components have the same size. 
In the $\delta$-function approximation of the synchrotron emissivity for an electron, the IC 
($E$ / 1 TeV) and X-ray ($\epsilon$ / 1 keV) synchrotron energies in Eq.~(\ref{eq:ICandXray}) are coupled according to 
(see Aharonian \etal \cite{Aharonian:4}):
\begin{equation}
\epsilon \sim 0.07 \left( \frac{E}{1\, \rm TeV} \right) \left( \frac{B}{10^{-5}\, \rm G}\right)  \rm keV
\end{equation}
so that for IC fluxes at $\sim$1 TeV, a comparison with the X-ray flux at energies $\sim$0.1 to 1.0 keV is required for  
values of $B \sim 10$ to 100 $\mu$G expected in a SNR. 
Eq.~(\ref{eq:ICandXray}) follows from the fact that the ratio of the synchrotron and IC energy fluxes is proportional to the ratio 
of the energy densities in the magnetic and CMB fields respectively. The dominant IC flux arises from the up-scattering of CMB photons 
by CR electrons of energy 10 to 100 TeV.
Preliminary results from RXTE data (Petre \etal \cite{Petre:1}) indicate a normalisation (at 1 keV) and photon index 
respectively of 0.30$\pm 0.02$ photons cm$^{-2}$ s$^{-1}$ and $-$3.18$\pm 0.02$, valid at energies from 10 to
20 keV. From earlier GINGA data Fink \etal (\cite{Fink:1}) derived a power law fit (in combination with a thermal bremsstrahlung and 
Fe line component) of photon index $-$2.72 (see Fink for error details) and normalisation of 7.4$^{+8.9}_{-7.4} \times 10^{-2}$ ph 
cm$^{-2}$s$^{-1}$ (at 1 keV), which is valid from 4.5 to 20 keV.  
A direct extrapolation to energies required for the $B$ field estimate, namely 0.1 to 1.0 keV, will lead to overestimation of the synchrotron 
flux since one would expect a turnover near this range. A more reasonable method to 
estimate fluxes at such lower energies is to fit to the X-ray flux a power law spectrum with exponential cutoff.
The RXTE data is preferred for such a fit due to the much reduced errors compared to those of GINGA data.
We fit the spectrum using the function:
\begin{equation}
 E^2\,F(E) = A E^{1-\alpha} \exp[-(E/E_{\rm m})^{0.5}] \,\,\,\, {\rm erg\, cm^{-2}\, s^{-1}}
 \label{eq:xrayfit}
\end{equation}
where $\alpha$ (radio photon index) and $E_{\rm m}$ (cutoff energy) are free parameters and $A$ is the normalisation determined from the 
flux density at 1 GHz $S_{\rm 1\, GHz}$=56 Jy (Green \cite{Green:1}). Eq.~(\ref{eq:xrayfit}) is derived for an electron spectrum with 
exponential cutoff using the $\delta$-function approximation for the electron synchrotron emissivity and is shown to adequately
describe the X-ray spectrum out to energies $\sim10E_{\rm m}$ (Reynolds \cite{Reynolds:1}). Our fit yields values of 
$\alpha$=0.7 and $E_{\rm m}$=1.6 keV respectively with $\alpha$ differing somewhat from the radio photon index, listed as 0.61 in Green's 
catalogue. Katz-Stone \etal (\cite{Katz-Stone:1}) summarise historical measurements of the radio index as varying between 0.5 and 0.7 
at different radio bands. From their VLA study, Katz-Stone indicate an average index of the filaments at 0.52.
Results are plotted in Fig.~\ref{fig:fits}a showing our best fit, and that when fixing $\alpha$=0.61. 
We note that Hwang \etal (\cite{Hwang:2}) have placed an upper limit to the non-thermal luminosity in the 
0.5 to 10 keV regime on the order of $\sim$10$^{34}$ erg s$^{-1}$ (0.5 to 10 keV), based on careful modelling of the non-thermal
ASCA detection. In order to accommodate their upper limit to the non-thermal luminosity (for a distance of 2.3 kpc 
as used by Hwang) our synchrotron spectrum fit to RXTE data must be scaled down by a factor $\sim$15.5, assuming 
that their upper limit is {\em exactly} 10$^{34}$ erg s$^{-1}$ (dotted line in Fig.~\ref{fig:fits}a). 

If we take the estimated synchrotron fluxes from 0.1 to 1 keV from the direct RXTE fit with our upper limit at 1 TeV for the IC flux, 
a range of $B_{-5}$ values 22 to 24 $\mu$G, plotted in Fig.~\ref{fig:fits}b (solid line), are obtained, which are reasonably
 consistent with that expected after amplification by the SNR shock. Using however the lower synchrotron 
energetics implied by Hwang \etal (\cite{Hwang:2}) we arrive at more conservative lower limits on $B_{-5}$ at
$\sim$6 $\mu$G, (dotted line). Such a value would result if little or no amplification was present. Allowing for
uncertainty of one order of magnitude in the Hwang estimate (ie. 5$\times10^{-33}$ to 4$\times10^{-34}$ erg s$^{-1}$) would,
for the upper bound case, give a $B$ field lower limit a factor $\sqrt{4}$ higher, ie. $\sim$12 $\mu$G. 

\section{Comparisons with DAV}
We will consider here the production of $\gamma$-radiation by the hadronic or $\pi^\circ$-decay channel described in
two-fluid model by Drury, Aharonian \& V\"olk (\cite{Drury:1}), hereafter DAV, that has been used extensively in the past (see also
Naito \& Takahara \cite{Naito:1}). The DAV prediction may be scaled simply according to 
the energy budget of the SNR $E_{\rm sn}$, density of the upstream target matter $n$, and distance to the remnant $d$:
\begin{equation}
  F_\gamma \propto \Theta \left( \frac{E_{\rm sn}}{10^{51} \rm \,erg}\right) \left(\frac{n}{1 \rm \, cm^{-3}}\right) \left(\frac{d}{1 \, \rm kpc}\right)^{-2}
 \label{eq:fluxscaling}
\end{equation}
The parameter $\Theta$ represents the fraction of energy available for the acceleration of CR that
are ultimately released into the ISM. The mean value of $\Theta$ per SNR in the galaxy is 
estimated in the range 0.05 -- 0.3 (Berezhko \& V\"olk \cite{Berezhko:5}),  and pertains to an evolutionary state well into
the Sedov phase when accelerated particles start to be released into the ISM. We shall 
adopt the reasonable, and often used value of $\Theta = 0.1$, particle spectral index $-$1.1 (nominal theory), and for purposes
of argument that Tycho has progressed well into the Sedov phase.

\begin{figure}
 \vspace{9cm}
 \includegraphics{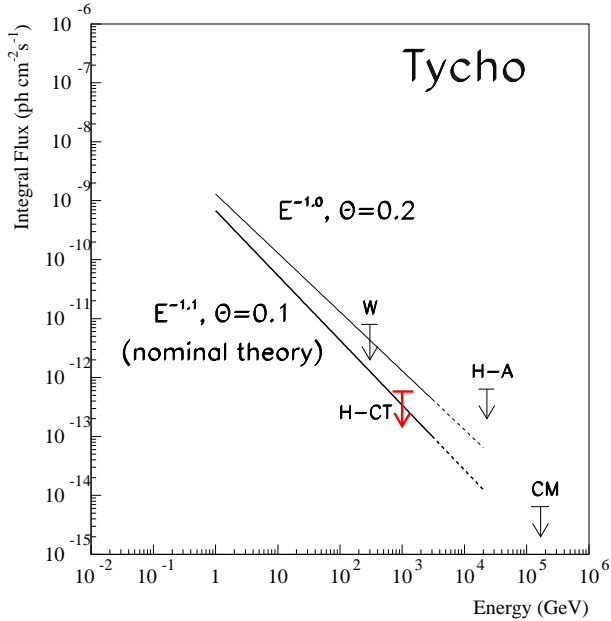}
 \caption{Upper limits from the present work (H-CT), and previous measurements (W - Whipple, H-A - HEGRA AIROBICC, CM - CASA-MIA. 
          See text for references)
          with predictions of the $\pi^\circ$-decay $\gamma$-ray flux from the DAV model. These DAV predictions from V\"olk \cite{Volk:1},
          use two limiting choices of parameters for $\Theta$, and the spectral index (indicated on the plot). Values for the other 
          parameters were
          $E_{\rm sn}=2\times 10^{50}$ erg (Smith \etal \cite{Smith:1}), $n$=1.0 cm$^{-3}$ and $d$=2.3 kpc (Heavens \cite{Heavens:1}). 
          Note that a reduction of the 'nominal theory' curve with a value $\theta=0.05$ may be argued in on the grounds that Tycho is 
          pre-Sedov (see text).}
 \label{fig:tycho_DAV}
\end{figure}

Fig.~\ref{fig:tycho_DAV} highlights a comparison of this DAV prediction under limiting parameter-space selections as calculated by
V\"olk (\cite{Volk:1}) with our present UL, and those of previous measurements
(Whipple, Buckley \etal \cite{Buckley:1} \& HEGRA AIROBICC, Prahl \etal \cite{Prahl:1}, and CASA-MIA, Borione \etal \cite{Borione:1}).

It is clear that with a factor $\sim$4 reduction between the Whipple and present upper limits (assuming a spectral index of $-$1.1
for this comparison), the DAV model is now constrained when allowing for a reasonable range of input parameters. We
are approaching the conservative boundary assumed by V\"olk (\cite{Volk:1}) for the Sedov phase. 

An important parameter in view of the expected TeV emission from a SNR is it's evolutionary phase. Generally accepted is 
the notion that the maximum TeV $\gamma$-ray emission occurs at the beginning of the Sedov phase and a broad maximum in the total CR 
energy is reached during the Sedov phase. At the former time the mass of swept-up material will just exceed that ejected, and the outer 
SNR shock begins to expand with 
radius $R\sim t^{2/5}$. Typical ages of Galactic SNRs for the onset of the Sedov phase are 500 to 1000 yrs. Tycho is a rather young SNR (428 yrs old) 
expanding into an ISM with density typical of that expected within the Galactic plane and so it may not be clear as to exactly what phase Tycho is 
presently in. 
Observational evidence indicates that globally, Tycho's SNR is near to the Sedov phase ($R\sim t^\nu$, $\nu=0.46 \pm0.02$, 
Tan \& Gull \cite{Tan:1}),
and is expanding into, on a global scale at least, a homogeneous ISM. Reynoso \etal (\cite{Reynoso:2}) report a similar value, 
from VLA data for the global expansion, but note that $\nu$ varies between 0.2 and 0.8 around the shell
and suggest the presence of
denser material along the eastern side (see also recent 21cm observations by Reynoso \etal (\cite{Reynoso:1})). The expansion rate
at X-ray energies is significantly higher than at radio energies, but as in the radio, it also varies considerably azimuthally
(Hughes \cite{Hughes:1}). In contrast to the radio behaviour however, the expansion rate at X-ray energies varies radially,
supporting the idea of an ejecta-dominated, or pre-Sedov evolutionary state.
Taking these results together with the Rayleigh-Taylor 
instabilities along the eastern side suggested by Velazquez \etal (\cite{Velazquez:1}), we may conclude that the global evolution state 
of Tycho's SNR is pre-Sedov, although at regions of high density ISM, the evolutionary state will be locally advanced. 

A comparison with DAV predictions should therefore allow for the current  
evolutionary state of Tycho. The lower curve of Fig.~\ref{fig:tycho_DAV} 
(nominal theory) may be lowered further
since the age of Tycho is likely less than the sweep-up time ($t_0$=555 yrs) which signals the
onset of the Sedov phase. A reduction in the $\gamma$-ray emissivity for pre-Sedov epochs can be
manifested as a reduction in the $\theta$ parameter to values $<$0.1. A quantitative estimate for $\theta$
in this case is not trivial but generally one can expect a reduction by a factor $\sim$2 based on consideration of results
in figs. 1 to 3 of DAV, describing different rates of luminosity increase with time.
In the next section comparisons to a model dealing with time-dependence in detail are made.

\section{Comparisons with Kinetic Theory}
The most recent nonlinear kinetic models are the time-dependent, spherically symmetric
solution of Berezhko \& V\"olk \cite{Berezhko:1}, hereafter BV, and the Monte Carlo simulation of a quasistationary outer
SNR shock in plane parallel geometry by Baring \etal \cite{Baring:3}. The BV model is 
based on the numerical solution of Berezhko, Yelshin \& Ksenofontov (\cite{Berezhko:4}) and invokes a 
distribution of ejecta velocities (e.g. Chevalier \& Liang \cite{Chevalier:1} and references therein) that contains very high speed 
components compared to the mean ejecta velocity. At early evolutionary phases this leads to much higher 
shock speeds than implied by the mean ejecta velocity, and thus to much more intensive CR and $\gamma$-ray
production. We will in this paper make use of BV as it was calculated in their paper, scaling the parameters to those of Tycho's SNR.
As we shall note shortly, in addition, a {\em physical} re-normalisation of these results is necessary. 
For a more complete treatment, 
we refer to a companion paper elsewhere in which the BV model will be calculated assuming Tycho parameters.

A scaling of the original BV calculations is required since the start-up parameters  
(in particular ejected mass $M_{\rm ej}=10M_\odot$ and SNR energy $E_{\rm sn}$=10$^{51}$ erg) may differ from those 
one might expect for Tycho's SNR. For this scaling
we may reasonably assume, that the ejecta mass only effects the initial
normalisation in the form of the sweep-up time $t_0$, and that other parameters will scale the flux
in the same way they do the DAV prediction, ie. independent of time. 
The first row of Table~\ref{tab:BVscaling} presents a numerical comparison of our upper limit with 
the predictions of the BV model, $F_{-13}^\gamma$ (in units 
of 10$^{-13}$ ph cm$^{-2}$ s$^{-1}$), scaled to Tycho parameters.
$\Theta_s$ is the released fraction of energy available for CRs and $\Theta=0.1$ is the value assumed for an average SNR, discussed previously.
\begin{table}
 \begin{center}
 \begin{tabular}{lrrrr} \\ \hline \hline
  Case                    &  a      &  b     &  c    &  d$^\dagger$   \\ \hline
  $B$ ($\mu$G)            &  5      & 30     &  5    &  5   \\
  $\eta$                  & 10$^{-3}$ & 10$^{-3}$ & 10$^{-4}$ & 10$^{-4}$ \\ \hline
  $F_{-13}^\gamma$        &  91.00  & 49.20  & 20.00 & 9.00  \\ 
  $\Theta_{\rm s}$              &  0.60   & 0.55   & 0.50  & 0.50 \\
  $\Theta/\Theta_{\rm s}$       &  0.17   & 0.18   & 0.20  & 0.20 \\
  $F_{-13}^{\gamma,{\rm r}}$    &  15.50  & 9.00   & 4.00  & 1.80 \\
  $F_{-13}^{\gamma,{\rm r}}$/UL &  2.60   & 1.50   & 0.69  & 0.31 \\
  $\Theta_{\rm r}(t=428)$       &  0.06   & 0.04   & 0.04  & 0.04 \\ \hline \hline
  \multicolumn{5}{l}{\scriptsize HEGRA CT-System UL = 5.78 (10$^{-13}$ ph cm$^{-2}$ s$^{-1}$)} \\
  \multicolumn{5}{l}{\scriptsize $\dagger$ Using a mean value for ejecta velocities.}
  \end{tabular}
 \end{center}
 \caption{Theoretical $\gamma$-ray fluxes, $F_{-13}^\gamma$ (10$^{-13}$ ph cm$^{-2}$ s$^{-1}$) and
          released relative CR energies $\Theta_{\rm s}$ from the BV model 
          scaled according to parameters of Tycho's SNR (see text). Each case represents a choice
          of injection rate $\eta$ and $B$-field (labeled in the caption of Fig.~\protect\ref{fig:modelcomp}).
          The parameters subscripted with $r$ refer to those additionally re-normalised to account for the effect of
          assuming in the BV model, a shock normal parallel to the B field (see results in Fig.\ref{fig:modelcomp}). The
          re-normalisation factor here is ($\Theta=0.1)/\Theta_{\rm s}$, and $\Theta_{\rm r}(t=428)$ is the relative CR energy
          calculated at the current age of Tycho's SNR.}
 \label{tab:BVscaling}
\end{table}
Inspection of $F_{-13}^\gamma$ for all cases in Table~\ref{tab:BVscaling} reveals that they 
are inconsistent with our upper limit. 

\begin{figure}
\vspace{6cm}
\includegraphics{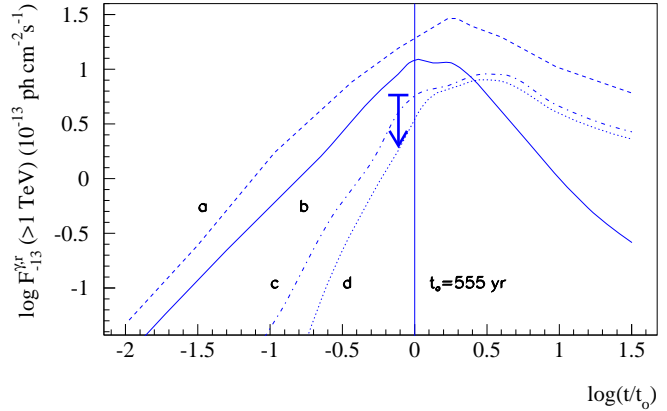}
\caption{Time dependence of the $\pi^\circ$ decay TeV $\gamma$-ray integral flux $F^{\gamma,r}_{-13}$ (10$^{-13}$ cm$^{-2}$ s$^{-1}$)
         from Tycho for
         an injection rate $\eta=10^{-3}$ and $B=5\mu$G (dashed line, case a), $\eta=10^{-3}$ and
         $B=30\mu$G (solid line, case b), $\eta=10^{-4}$ and $B=5\mu$G (dash-dotted line, case c).
         The dotted line (case d) represents a less realistic single velocity ejecta case with $\eta=10^{-4}$ and $B=5\mu$G,
         as is adopted in the DAV model.
         $t_0$ is the time of Sedov phase onset. These results have been scaled according to parameters for
         Tycho ($M_{\rm ej}=1.4M_\odot$, $E_{\rm sn}=2\times10^{50}$ erg, $n=1.0$ cm$^{-3}$, $d$=2.3 kpc. $t_0$ is estimated at 555 yrs, 
         somewhat larger than
         the current age of 428 yrs), and re-normalised by the ratio $\Theta/\Theta_{\rm s}$. Original model 
         calculations are
         from Berezhko \& V\"olk (\cite{Berezhko:1}).}
\label{fig:modelcomp}
\end{figure}

However, we must now invoke a further, physical re-normalisation of the Tycho-scaled BV
predictions for the following reason. The BV model assumes a parallel shock geometry (shock normal parallel
to the average up-stream $B$ field) over the entire spherical shock surface. For spherical geometry expected of a SNR, such
an assumption will only apply for a limited part of the SNR shock surface. Over other parts the shock is quasi-perpendicular.
A strongly reduced injection efficiency below that derived from injection 
theory for parallel shocks ($\eta\sim 10^{-3}-10^{-4}$) along with a commensurate reduction of the particle production will 
be noticed in such regions
(regarding the question of injection in general, see Kirk \& Dendy \cite{Kirk:1} for a recent review). 
Correcting for this effect in detail is complicated, and will be discussed in a follow-up paper. However, a rough, but empirically-argued
implementation of a re-normalisation is achieved by reducing the fluxes $F^{13}_\gamma$ by the ratio
$\Theta/\Theta_{\rm s}$ to a re-normalised flux $F_{-13}^{\gamma,{\rm r}} = F_{-13}^{\gamma}\Theta/\Theta_{\rm s}$, where $\Theta$=0.1 is 
the empirically 
expected value for an average Galactic SNR discussed earlier (and adopted by DAV), and $\Theta_{\rm s}$ is that value predicted by the BV model. 
The re-normalised fluxes are those $r$-subscripted in 
Table~\ref{tab:BVscaling}, and are rather close to our upper limit and the relative CR energy calculated
at the present age of Tycho. $\Theta_{\rm r}(t)=E_{\rm cr}(t)/E_{\rm sn} \times \Theta/\Theta_{\rm s}$ corresponds to values
of around $10^{49}$ ergs. Fig.~\ref{fig:modelcomp} graphically compares 
$F^{\gamma,r}_{-13}$ for the different cases discussed. It appears that cases (c) 
and (d), with a lower injection rate and $B$-field, are preferred. However, the combinations of $B$ and $\eta$ used here are not 
exhaustive. For example a case assuming higher $B$ (perhaps suggested by our inverse Compton interpretation in section~\ref{sec:ICint})
and low injection rate has not been tested by BV. More thorough comparisons will be made later. The uncertainties in 
scaling parameters, $d$, $E_{\rm sn}$ and $n$ will weaken of course the conclusion favouring cases (c) and (d), in which lower injection rates
and $B$ fields are assumed.
The uncertainty in $d$, where roughly a factor of two above the 
value used here (2.3 kpc) is published (Schwarz \etal \cite{Schwarz:1}), will have strong influence. Also, a somewhat lower value of $n$=0.3
cm$^{-3}$  derived by Seward \etal (\cite{Seward:1}) for the pre-shock density may also be preferable to the value given by Smith \etal (\cite{Smith:1}) used here.  
At this point we would conclude that a general consistency with the BV model is obtained, perhaps favouring lower injection rates and
higher $B$ fields (noting the reduction in emissivity between cases (a) and (b), due to a change in $B$ field), but 
note that use of a wider parameter space under a dedicated BV calculation is required.

\section{Conclusion}
A search for TeV $\gamma$-radiation from Tycho's SNR has been performed over two years (1997 \& 1998) with the HEGRA 
CT-system. We find no evidence for such emission
and the upper limit (3$\sigma$ level) to the TeV flux is estimated at 33 milli-Crab or $5.78\times 10^{-13}$ ph cm$^{-2}$s$^{-1}$, 
a value which is a factor $\sim$4
less than that previously published by the Whipple collaboration (when assuming a spectral index of $-$1.1 in
for the comparison). 

Making use of preliminary RXTE results, it is possible to set a lower limit on the magnetic
field in Tycho $B\geq22$ $\mu$G under the assumption that the observed hard or non-thermal tail in the X-ray spectrum is  
attributed to synchrotron radiation from multi-TeV electrons.  A more conservative $B$ field lower limit of $\sim$6 $\mu$G is obtained 
if we use the upper limit to the non-thermal
X-ray flux estimated by Hwang \etal (\cite{Hwang:2}).
It may therefore be premature to draw conclusions about the minimum magnetic field until full analysis of RXTE data
are complete. Interestingly the upper limit to the non-thermal energetics of Hwang \etal (\cite{Hwang:2}) is well below that implied by a
purely non-thermal interpretation of the RXTE and GINGA results.
 
Comparisons to a long-adopted model of TeV emission from the decay of $\pi^\circ$ 
were made (Drury, Aharonian \& V\"olk \cite{Drury:1}, denoted DAV), in the initial case assuming that Tycho's SNR is well in to the 
Sedov phase of evolution. Our upper limit is close to a conservative prediction of DAV.
Uncertainty in the global evolutionary state (Sedov, or pre Sedov) 
of Tycho presents a complication upon comparison to the DAV $\pi^\circ$ decay $\gamma$-ray predictions. 
As a function of time, advanced kinetic theory (Berezhko \& V\"olk \cite{Berezhko:1}, denoted BV) indicates a relatively early
rise in the $\gamma$-ray emissivity of the hadronic channel for pre-Sedov epochs, in contrast to the less 
realistic case of single-velocity ejecta, as adopted in the DAV model. We find that
after adjusting the predictions of BV, firstly scaling for Tycho's parameters and secondly, re-normalising to account for 
quasi-perpendicular shock directions 
expected in an SNR, a general consistency with our upper limit is found for a range of injection rates and $B$ fields.

We conclude from this non-detection and the interpretation assuming Tycho to be a source of multi-TeV
electrons and hadrons that Tycho's SNR is yet to be ruled out as an average accelerator of Galactic CRs. Our upper limit 
supports the notion that Tycho is in a pre-Sedov evolutionary state.
A more complete investigation will be left to a companion paper in which a dedicated
calculation in the framework of the BV model will be performed for the parameters (and their uncertainties) of Tycho's SNR.

Resulting from a type Ia supernova, Tycho's SNR in principle represents
the simplest category in terms of a theoretical understanding of the SNR dynamics and interaction with the 
ISM. Yet, given the complexity in modeling non-thermal processes, the basic consistency between 
experiment and theoretical predictions certainly encourages future observations of Tycho's SNR and other examples of it's
class. 
Because of the $\sqrt{t}$ dependence on instrument sensitivity, it is likely that further observations 
at TeV energies of Tycho will have to wait for the 
next generation of telescopes operating in the northern hemisphere (e.g MAGIC, Lorenz \etal \cite{Lorenz:1}; VERITAS, 
Krennrich \etal \cite{Krennrich:1}). With roughly one order of magnitude 
improvement in sensitivity for the next generation instruments over
that currently available, and the fact that current theory does not leave much room for non-detection at higher 
instrument sensitivity, a decisive test of the question of whether or not Tycho's SNR 
contributes to the Galactic CR population at an average level is expected within the next few years.

 \begin{acknowledgements}
We acknowledge the financial support of the German Ministry for Research and Technology (BMBF) and the Spanish
Research Council. We thank the Instituto Astrof\'{i}sica de Canarias for the use of their site and for
the excellent working conditions at La Palma. The technical support staff of the Heidelberg, Kiel, Munich
and Yerevan Institutes are also acknowledged. Glenn Allen is thanked for providing a fit to the RXTE Tycho data, 
and Anne Decourchelle for the model fit to the XMM data. GPR gratefully acknowledges receipt of a von Humboldt Fellowship.
An anonymous referee is thanked for helpful comments.
 \end{acknowledgements}


\begin{thebibliography}{xx}

  \bibitem[1997]{Aharonian:4} Aharonian F.A., Atoyan A.M., Kifune T., 1997, MNRAS 291, ~162
  \bibitem[1999]{Aharonian:6} Aharonian F.A. \& Atoyan A.M., 1999, A\&A 351, ~330
  \bibitem[2000a]{Aharonian:1} Aharonian F.A, Akhperjanian A.G., Andronache M. et~al. 2000a, A\&A, {\em submitted} 
  \bibitem[2000b]{Aharonian:2} Aharonian F.A, Akhperjanian A.G., Barrio J.A. et~al. 2000b, ApJ 539, ~317 
  \bibitem[2000c]{Aharonian:3} Aharonian F.A., Akhperjanian A.G., Barrio J.A. et~al. 2000c, A\&A 353, ~847 
  \bibitem[2000d]{Aharonian:5} Aharonian F.A., Akhperjanian A.G., Barrio J.A. et~al. 2000d, A\&A 361, ~1073 
  \bibitem[1986]{Albinson:1} Albinson J.S. et~al. 1986, MNRAS 219, ~427
  \bibitem[1995]{Allen:1} Allen G.E., Berley D., Biller S., et~al. 1995, ApJ 448, ~L25
  \bibitem[1990]{Asvarov:1} Asvarov A.I., Dogiel V.A., Guseinov O.H., Kasumov F.K. 1990, A\&A 229, ~196
  \bibitem[2000]{Atoyan:1} Atoyan A.M., Aharonian F.A., Tuffs R.J., V\"olk H.J.  2000, A\&A 355, ~211
  \bibitem[2000a]{Baring:1} Baring M.G. 2000a, {\em in Rapportuer Vol.} 26th ICRC, ed. B.L. Dingus (AIP, New York, 2000)
  \bibitem[2000b]{Baring:2} Baring M.G. 2000b, {\em Towards a Major Atmospheric Cherenkov Detector - VI} ed. B.L. Dingus, M.H. Salamon,
                             D.B. Kieda (AIP Conf. Series 515, New York, 2000)
  \bibitem[1999]{Baring:3} Baring M.G., Ellison, D.C., Reynolds, S.P. \etal 1999, ApJ 513, ~311
  \bibitem[1998]{Barrio:1} Barrio, J.A. et~al. 1998, ``The MAGIC Telescope'', design study, MPI-PhE/98-5
  \bibitem[1994]{Berezhko:4} Berezhko, E.G., Yelshin V.K., Ksenofontov L.T. 1994, Astropart. Phys. 2, ~215
  \bibitem[1997]{Berezhko:1} Berezhko E.G. and V\"olk H.J. 1997, Astropart. Phys. 7, ~183 
  \bibitem[1999]{Berezhko:3} Berezhko, E.G, Ksenofontov L.T., Petukhov S.I. 1999, Proc. 26th Int. Cosmic Ray Conf. (Salt Lake City) 4, ~431
  \bibitem[2000]{Berezhko:5} Berezhko, E.G. and V\"olk H.J. 2000, ApJ 540, ~923
  \bibitem[1995]{Borione:1} Borione A., Catanese M., Covault C.E. \etal 1995, Proc. 24th Int. Cosmic Ray Conf. (Rome) 2, ~439
  \bibitem[1998]{Buckley:1} Buckley J.H., Akerlof C.W., Carter-Lewis D.A., et~al. 1998, A\&A 329, ~639
  \bibitem[1989]{Chevalier:1} Chevalier R.A. and Liang E.P. 1989, ApJ 344, ~332
  \bibitem[2001]{Decourchelle:1} Decourchelle A., Sauvageot J.L., Audard M. \etal 2001, A\&A 365, ~L218
  \bibitem[1994]{Drury:1} Drury L. O'C, Aharonian F., V\"{o}lk H.J. 1994, A\&A 287, ~959
  \bibitem[1994]{Fink:1} Fink H.H., Asaoka I., Brinkman W. et~al 1994, A\&A 283, ~635
  \bibitem[1999]{Goret:1} Goret P., Gouiffes C., Nuss E., et~al. 1999, Proc. 26th Int. Cosmic Ray Conf. (Salt Lake City) 3, ~496
  \bibitem[2000]{Green:1} Green D.A., 2000, `A Catalogue of Galactic Supernova Remnants (2000 August version)', Mullard Radio Astronomy 
                            Observatory, Cavendish Laboratory, Cambridge, United Kingdom (available on the World-Wide-Web at
                              {\tt http://www.mrao.cam.ac.uk/surveys/snrs/}). 
  \bibitem[1984]{Heavens:1} Heavens A.F. 1984, MNRAS 211, ~195
  \bibitem[1983]{Helene:1} Helene O. 1983, Nucl. Inst. Meth. 212, ~319
  \bibitem[1999]{Hofmann:1} Hofmann W., Jung I., Konopelko A. et~al. 1999, Astropart. Phys. 12, ~135
  \bibitem[1997]{Hwang:1} Hwang U., Gotthelf E.V. 1997, ApJ 475, ~665
  \bibitem[1998]{Hwang:2} Hwang U., Hughes J.P., Petre R. 1998, ApJ 497, ~833
  \bibitem[2000]{Hughes:1} Hughes J.P. 2000, ApJ, {\em in press}, {\tt astro-ph/0010122}
  \bibitem[2000]{Katz-Stone:1} Katz-Stone D.M., Kassim N.E., Joseph T., et~al 2000, AJ 529, ~453
  \bibitem[2001]{Kirk:1} Kirk J.G., Dendy R.O. 2000, J.Phys. G: Nucl.Part.Phys. {\em in press}
  \bibitem[1999]{Konopelko:1} Konopelko A., Hemberger M., Aharonian F.A., et~al. 1999, Astropart. Phys. 10, ~275
  \bibitem[1995]{Koyama:1} Koyama K., Petre R., Goffhelf E.V. et~al. 1995, Nat 378, ~255
  \bibitem[1997]{Koyama:2} Koyama K., Kinusaga K., Matsuzaki K. et~al. 1997, PASJ 49, ~L7
  \bibitem[1999]{Krennrich:1} Krennrich F. \etal 1999, {\em Towards a Major Atmospheric Cherenkov Detector - VI}, pg 515, ed B.L. Dingus, 
                                    M.H. Salamon, D.B. Kieda (AIP Conf. Series 515, New York, 2000) 
  \bibitem[1998]{Laming:1} Laming J.M. ApJ 499, ~309
  \bibitem[1983]{Li:1} Li T., Ma Y. 1983, ApJ 272, ~317
  \bibitem[1997]{Longair:1} Longair M.S. {\em High Energy Astrophysics} (Cambridge University Press, 1997)
  \bibitem[1999]{Lorenz:1} Lorenz E. \etal 1999, {\em Towards a Major Atmospheric Cherenkov Detector - VI}, pg 510, ed B.L. Dingus, 
                                    M.H. Salamon, D.B. Kieda (AIP Conf. Series 515, New York, 2000) 
  \bibitem[2000]{Muraishi:1} Muraishi H., Tanimori T., Yanagita S., et~al. 2000, A\&A 354, L57
  \bibitem[1994]{Naito:1} Naito T., Takahara F. 1994, J Phys G: Nucl Part Phys 20, ~477
  \bibitem[1999]{Petre:1} Petre R., Allen G.E., Hwang U. 1999, Astronomische Nachrichten, 320, ~199
  \bibitem[1997]{Prahl:1} Prahl J., Prosch C. for the HEGRA collab. 1997, Proc. 25th Int. Cosmic Ray Conf. (Durban) 3, ~217 
  \bibitem[1996]{Prosch:1} Prosch C., Feigl E., Plaga, R., et~al. 1996, A\&A 314, ~275
  \bibitem[1999]{Puhlhofer:1} P\"{u}hlhofer G., V\"{o}lk H., Wiedner C.A., et~al. 1999, 
                 Proc. 26th Int. Cosmic Ray Conf. (Salt Lake City) 3, ~492 
  \bibitem[1999]{Reynolds:1} Reynolds S.P., Keohane J.W. 1999, ApJ 525, ~368  
  \bibitem[1999]{Reynoso:1} Reynoso E.M., Velazquez P.F., Dubner G.M., Goss W.M. 1999, AJ 117, ~1827
  \bibitem[1997]{Reynoso:2} Reynoso E.M., Moffett D.A., Goss W.M. \etal 1997, ApJ 491, ~816
  \bibitem[2000]{Rowell:1} Rowell G.P., Naito T., Dazeley S.A., et~al. 2000, A\&A 359, ~337 
  \bibitem[1995]{Schwarz:1} Schwarz U.J., Goss W.M., Kalberla P.M., Benaglia P. 1995 A\&A 299, ~193
  \bibitem[1983]{Seward:1} Seward F., Gorenstein P., Tucker W. 1983, ApJ 266, ~287
  \bibitem[1988]{Smith:1} Smith A., Davelaar J., Peacock A., et~al. 1988, ApJ 325, ~288
  \bibitem[1985]{Tan:1} Tan S.M. and Gull S.F. 1985, MNRAS 216, ~949
  \bibitem[1998]{Tanimori:1} Tanimori T., Hayami Y., Kamei S. et~al. 1998, ApJ 492, ~L33 
  \bibitem[1998]{Tatischeff:1} Tatischeff V., Ramaty R., Kozlovsky B. 1998, ApJ 504, ~874
  \bibitem[1998]{Velazquez:1} Velazquez P.F., Gomez D.O., Dubner G.M. \etal 1998, A\&A 334, ~1060
  \bibitem[1997]{Volk:1} V\"olk H.J. 1997 {\em in ``Towards a Major Atmopsheric Cerenkov Detector - V''}, ed O.C. deJager
                                      (WESPRINT, Potchefstroom 1997), ~87
 \end{thebibliography}
\end{document}